# How oxygen influences the catalytic activity of iron during carbon nanotube nucleation

Ben McLean,[1,2]* Alister J. Page,[3] Feng Ding[1,4]

[1] Center for Multidimensional Carbon Materials (CMCM), Institute for Basic Science (IBS), Ulsan 44919, Republic of Korea

[2] School of Engineering, RMIT University, Victoria, 3001, Australia

[3] Discipline of Chemistry, The University of Newcastle, Callaghan, New South Wales 2308, Australia

[4] Suzhou Laboratory, Suzhou 215100, China.

**Abstract**

The catalytic activity of metal nanoparticles toward nucleation of single-walled carbon nanotubes (SWCNTs) is fundamental to achieving structure-controlled growth using catalytic chemical vapor deposition (CVD). Despite the success of oxidized catalysts in SWCNT growth, there is a lack of understanding regarding how oxygen influences the catalysts and the nucleation process. Quantum chemical molecular dynamics (MD) simulations employing density functional tight binding (DFTB) demonstrate that the kinetics of carbon nucleation on an iron nanoparticle catalyst can be tuned via oxygen loading. Increasing the oxygen content in the catalyst leads to activation of surface-bound carbon species and enhanced carbon chain growth due to respective weakening and strengthening of the C-C and Fe-C bonding. This is due to oxygen modulating the electronic structure of the iron catalyst, with the Fermi level of the catalyst increasing proportionally with oxygen content until the iron:oxygen stoichiometry reaches parity. The increase in Fe 3d states near the Fermi level also promotes the donation of electron density into unoccupied C 2p states, activating C-C bonds which in turn facilitates carbon chain growth and slows carbon ring condensation.

## 1. Introduction

Following extensive research regarding catalytic chemical vapor deposition (CVD) synthesis of single-walled carbon nanotubes (SWCNTs), it is undeniable that the catalyst plays a crucial role in nucleation and growth.[1-3] Traditionally, pure transition metals, such as Fe, were effective as growth catalysts. However, in efforts to achieve control over *(n,m)* SWCNT chirality and diameter via 'catalyst design', alternatives such as FeMo,[4-7] FeCo[8, 9] and FeRu[10] alloy catalysts have been explored. Experimentally, $Fe_2O_3$ has previously been demonstrated as a prominent growth catalyst for long SWCNTs with controlled diameter,[11, 12] and the oxidation of the Fe catalyst is an important aspect of $H_2O$-assisted supergrowth.[13, 14] Molecular simulations have contributed significantly to understanding the growth of SWCNTs on Fe,[3, 15-22] though few studies have investigated growth on alternative catalysts. In one example, Fukuhara *et al.*[23] studied the dissociation of ethanol on a $Fe_{16}Co_{16}$ catalyst cluster using *ab initio* molecular dynamics (MD), demonstrating the benefits of unique reaction sites of an alloy catalyst for favoring select dissociation events e.g. C-C or C-O bond cleavage. Wang *et al.*[24] explored the role of oxygen in the nucleation of SWCNTs from $C_2$ over 330 ps on $Fe_{38}$, $Fe_{38}O_x$ (x=1-10) with density functional tight binding (DFTB) MD simulations. These simulations revealed that ring nucleation kinetics were modulated by the amount of oxygen in the catalyst nanoparticle; nucleation kinetics were accelerated for the $Fe_{38}O_6$ model catalyst, while slower nucleation kinetics were observed on $Fe_{38}O_{10}$. The presence of oxygen hindered the formation of larger clusters, despite primarily residing within the catalyst nanoparticle with little observable influence at the growth interface. MD simulations[25] of alcohol CVD on Fe demonstrated that the extent of catalyst oxidation and preferred reaction pathways are mediated by the OH radical.

From these studies, the precise role of oxygen during CVD growth of SWCNTs remains unresolved. Herein, we present quantum chemical MD simulations of FeO-catalyzed CVD, wherein the oxygen content within the Fe catalyst is varied. We explore how oxygen influences catalytic action in SWCNT nucleation, namely via the formation of growth precursors, such as branched carbon chains and SWCNT ring networks. Importantly, we examine how the degree of oxidation modulates the electronic structure of the catalyst, resulting in marked changes in the observed catalytic efficiency. On the basis of these results, we propose a new mechanism to explain oxygen's observed roles during FeO-catalyzed CVD SCWNT growth.

## 2. Methods

Self-consistent charge DFTB (SCC-DFTB) MD simulations[26] were performed in conjunction with the trans3d-0-1[27] and mio-1-1[26, 28] parameter sets, previously used to accurately simulate systems containing Fe, O, and C regarding SWCNT growth.[24, 29-31] Newton's equations of motion are integrated with the Velocity-Verlet algorithm[32] with a time step of 1.0 fs. An electronic temperature of $10^4$ K was used as validated in previous reports to ensure electronic convergence of Fe.[33, 34] All MD simulations employ canonical (NVT) ensembles and were held at 1273 K via the Nose-Hoover chain thermostat (chain length = 3, coupling strength = 3000 cm$^{-1}$).[35, 36] All MD simulations were performed using the DFTB+ software package.[37]

The nucleation and growth of SWCNTs on oxidized Fe catalysts is modelled in a cubic simulation box with side length 30 Å. A $Fe_{38}$ nanoparticle was selected as the catalyst for direct comparison with previous simulations, and oxidized by adding 10, 30 or 50 oxygen atoms – thus representing a range of loadings from oxygen-poor to oxygen saturation. Prior to growth, the oxidized $Fe_{38}$ nanoparticle is annealed for 100 ps at 1273 K. $C_2$ dimers are supplied to the simulation box at a rate of 1 $C_2$ with the same kinetic energy of $C_2$ at 1273 K, every 10 ps, 5 Å from the nanoparticle. Ten independent trajectories were observed such that representative mechanisms and quantitative data are obtained. All SWCNT nucleation data presented herein is an average across those ten trajectories. For each trajectory, after 30 $C_2$ dimers are supplied (300 ps elapsed), carbon supply ceases and the system is annealed at 1273 K for a further 500 ps. We distinguish these phases of the simulation as supply and annealing for clarity during discussion.

From these MD simulations, 5000 structures were selected at random from each simulation condition for Fermi level analysis, representing a 10% sampling density. For each of these structures, carbon atoms were removed from the system and the electronic structure was calculated using the SCC-DFTB method detailed above (without geometry optimization).

### 3. Results and Discussion

The mechanism of SWCNT growth itself is not our primary focus in this manuscript; the discussion below instead focuses on new insights into the nucleation of SWCNTs for oxidized Fe catalysts, and most importantly how the degree of catalyst oxidation influences the nucleation of carbon clusters on Fe catalysts. We therefore begin our discussion by briefly summarizing the main mechanistic steps towards SWCNT nucleation we observe in our CVD simulations, noting that they are consistent with prior simulations of SWCNT growth on pure Fe nanoparticles,[38-40] and oxidized Fe catalysts.[24, 25] The supply of $C_2$ dimers to the $Fe_{38}$ nanoparticle results initially in the surface diffusion and interaction of C via C-C bonds to nucleate carbon chains on the $Fe_{38}$ surface. These chains grow upon further $C_2$ addition and adjacent chains combine to form branched carbon clusters comprising y-shaped junctions. If the branched carbon chains are of sufficient length, they can oligomerize into pentagonal rings first, and subsequently hexagonal rings within a growing graphitic network. The discussion below is structured according to these principal mechanistic steps.

### 3.1 How oxygen influences carbon diffusion and chain formation

For each of the $Fe_{38}$, $Fe_{38}O_{10}$, $Fe_{38}O_{30}$ and $Fe_{38}O_{50}$ catalysts, Figure 1a plots the average carbon cluster size, Figure 1b plots the average carbon chain length and Figure 1c plots the number of C-C bonds observed as a function of time (data for all bond types are provided in Figure S1). As adjacent carbon chains bond, they oligomerize and form rings. The population of pentagonal, hexagonal and heptagonal rings formed over time for each catalyst are plotted in Figures 1d, 1e and 1f, respectively.

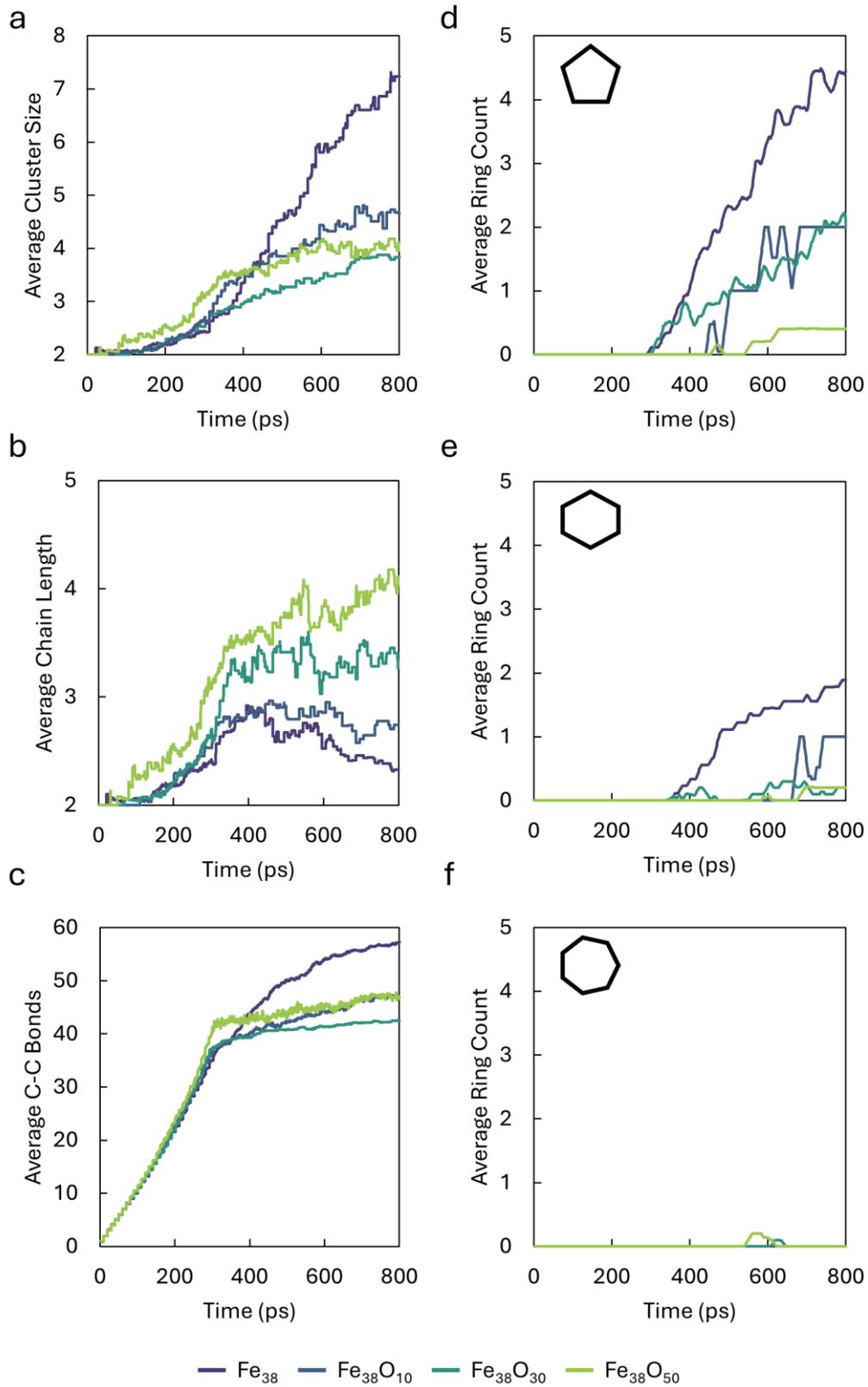

**Figure 1.** Average (a) carbon cluster size, (b) carbon chain lengths, (c) C-C bond counts, and populations of (d) pentagonal, (e) hexagonal and (f) heptagonal ring populations over 800 ps of simulation following the supply of 30 $C_2$ dimers to pure $Fe_{38}$ or $Fe_{38}$ containing either 10, 30, or 50 oxygen atoms. Bond counts for all bond types are presented in Figure S1. Data presented is averaged across ten unique trajectories.

During the supply of $C_2$ dimers to the catalyst surface, collision of $C_2$ on the surface occurs more frequently with an increased concentration of oxygen in the Fe catalyst, leading to the largest clusters on average present on $Fe_{38}O_{50}$, than $Fe_{38}O_{30}$, $Fe_{38}O_{10}$ and pure $Fe_{38}$, evident in Figure 1a. During annealing (after 300 ps) the rate of C clustering remains similar for $Fe_{38}O_{10}$ and $Fe_{38}O_{30}$ catalysts but slows for $Fe_{38}O_{50}$. Pure Fe exhibits the greatest rate of C clustering amongst the catalysts considered here. By 450 ps, the pure Fe catalyst on average hosts the largest C clusters, which on average are nearly double the size of those on the other catalysts after 800 ps of simulation. This is consistent with the process of growth proposed by Wang *et al.*[24] though more pronounced over the longer timescales investigated here and increase in oxygen loading of the catalyst. As $C_2$ is supplied to and diffuses across the catalyst surface, C-C bonds form as part of linear carbon chains on the surface of the catalyst, with longer average lengths over the course of the simulation on catalysts containing more oxygen, as seen in Figure 1b. In Figure 1c, we observe the increased rates of C-C bond formation with the inclusion of oxygen in the catalyst compared to pure Fe during supply. The rate of C-C bond formation and cluster formation is accelerated during annealing on pure Fe and $Fe_{38}O_{10}$, while the number of C-C bonds remains relatively constant for $Fe_{38}O_{30}$ and $Fe_{38}O_{50}$. The average length of the carbon chains after 300 ps continues to increase for $Fe_{38}O_{30}$ and $Fe_{38}O_{50}$ as the oligomerisation and combination of chains into clusters and rings is slower for more oxidized catalysts. For pure Fe, and for $Fe_{38}O_{10}$ the average chain length decreases, indicating the formation of rings, as discussed below. After 100 ps of annealing (400 ps total simulation time), rings are observed on pure Fe, while branched carbon junctions have formed on $Fe_{38}O_{10}$, and long chains dominate on highly oxidized catalysts, $Fe_{38}O_{30}$ and $Fe_{38}O_{50}$. Representative structures after 100 ps of annealing are shown in Figure S2.

To highlight the influence oxygen has on the Fe-C interaction, we plot (Figure S3a) the average partial atomic charges of each atom type over the CVD simulation. The presence of oxygen in the catalyst induces a strong positive charge on the Fe atoms, increased further by the adsorption of carbon. Carbon initially added to the surface has a partial atomic charge of approximately -0.25 |e| compared to the strong negative charge of oxygen (up to -0.77 |e|), allows it to diffuse quickly over the Fe surface as the Fe interacts more strongly with the O. With more oxygen in the catalyst, the slower clustering rate arises from a marked decrease in the average atomic charge of carbon, becoming more negative once more carbon is added to the surface. The average atomic charge of carbon is -0.31 |e|, -0.41 |e| and -0.57 |e| for 10, 30 and 50 oxygen atoms in the catalyst, respectively. This results in stronger Fe-C interactions and weaker C-C interactions on oxidized catalysts. The relatively weaker Fe-C interaction for pure Fe contributes to the enhanced C-C bonding and ring nucleation during annealing. Notably, this phenomenon was not observed by Wang et al.,[24] with no significant difference in the partial atomic charge for carbon between $Fe_{38}O_6$ and $Fe_{38}O_{10}$ catalysts, though revealed here by investigating catalysts with higher oxygen content.

Post-diffusion processes following $C_2$ supply such as chain oligomerisation to form rings and graphitic network formation are largely impeded by the presence of oxygen. Wang et al.[24] observed the nucleation of rings on oxidized catalysts earlier than pure $Fe_{38}$ under the same simulation conditions however we observe different behavior here, as discussed below. It is likely that the rapid supply of $C_2$ (1 $C_2$/ps) in their simulations led to faster kinetics of nucleation. Here, supplying 1 $C_2$/10 ps allows for diffusion of $C_2$ units on the catalyst surface to form longer carbon chains prior to ring nucleation. This leads to an important distinction regarding the nature of the carbon clusters on the oxidized Fe catalysts as demonstrated in Figure 1c. During supply, $C_2$ dimers primarily diffuse and collide to form linear carbon chains, hence the average chain length (Figure 1c) increases similarly to the average cluster size (Figure 1a). However, as these chains need to diffuse and oligomerize to nucleate ring structures and graphitic networks, there is a marked change during annealing. The presence of oxygen impedes the nucleation of ring structures and results in the larger clusters and chains observed on oxidized Fe catalysts until ~400 ps to remain as long chains throughout the simulation. The stronger Fe-C interaction evidenced by the average partial atomic charges described above (Figure S3a) diminishes the ability of larger chains structures to diffuse and rotate on the catalyst surface. Further, C-O bonding is not observed in these simulations during supply or annealing. On the pure Fe catalyst however, the nucleation of rings is unimpeded, and the cluster size increases significantly while the average chain length decreases. This is similar for $Fe_{38}O_{10}$ however the competition between $C_2$ diffusion to form chains and the closure of rings from these chains results in a more gradual increase in cluster size observed while the average chain length decreases slightly. This alludes to a more controlled growth process with the presence of some oxygen in the catalyst, but not so much to impede ring closure.[24]

Achieving the ideal balance of carbon diffusion and chain nucleation that can subsequently oligomerize into rings and larger graphitic networks is clearly dependent on the oxygen content of the catalyst. The addition of oxygen to the catalyst tunes the nature of the catalyst from solid to liquid-like under typical growth conditions. Figure S3b plots the average Lindemann Index of the Fe atoms in each catalyst over the 800 ps of simulation and shows the liquid-like nature of the pure Fe catalyst (Lindemann Index ~0.24) compared to the slightly liquid-like $Fe_{38}O_{10}$ (~0.12) and solid $Fe_{38}O_{30}$ and $Fe_{38}O_{50}$ (~0.07, ~0.06 respectively) nanoparticles. We discuss the influence of catalyst phase and oxygen content on the nucleation of rings below.

## 3.2 How does oxygen influence ring nucleation?

From Figure 1 and the discussion above, oxygen clearly enhances initial carbon chain and junction formation, but a high oxygen content prevents the closure of these chains into rings. On a pure Fe catalyst the carbon chains on the surface oligomerize into either pentagons or hexagons. The "pentagon-first" mechanism observed for pure Fe catalysts[38] is also observed on oxidized catalysts, as evident in Figure 2. For the oxidized $Fe_{38}O_{30}$ catalyst, the nucleation of pentagonal and hexagonal rings occurs after similar simulation times, though pentagons dominate. From our CVD simulations, it appears that surface or subsurface oxygen does not significantly influence the first ring nucleated. However, the presence of oxygen certainly influences ring nucleation and growth of ring networks. The representative ring nucleation mechanisms and the timescales they occur on as shown in Figure 2 are indicative of enhanced chain growth yet impeded ring condensation on catalysts with higher oxygen loading. Long chains are formed within 500 ps on $Fe_{38}O_{30}$ and $Fe_{38}O_{50}$ yet the closure of the first pentagon ring is significantly slower, particularly for $Fe_{38}O_{50}$. We observe that nucleated rings on oxidized catalyst surfaces remain bound to the same Fe atoms due to the strong Fe-C interaction, slowing the growth rates of these clusters and controlling the addition of C atoms (or $C_2$ dimers) to the cluster edge. We do not observe this behavior for rings nucleated on the pure Fe catalyst, which can diffuse more easily, or interact with multiple, more mobile Fe atoms in the catalyst. The enhanced mobility of Fe atoms in the pure Fe catalyst particles compared to those containing oxygen is indicated by the Lindemann Index for each particle plotted in Figure S3b.

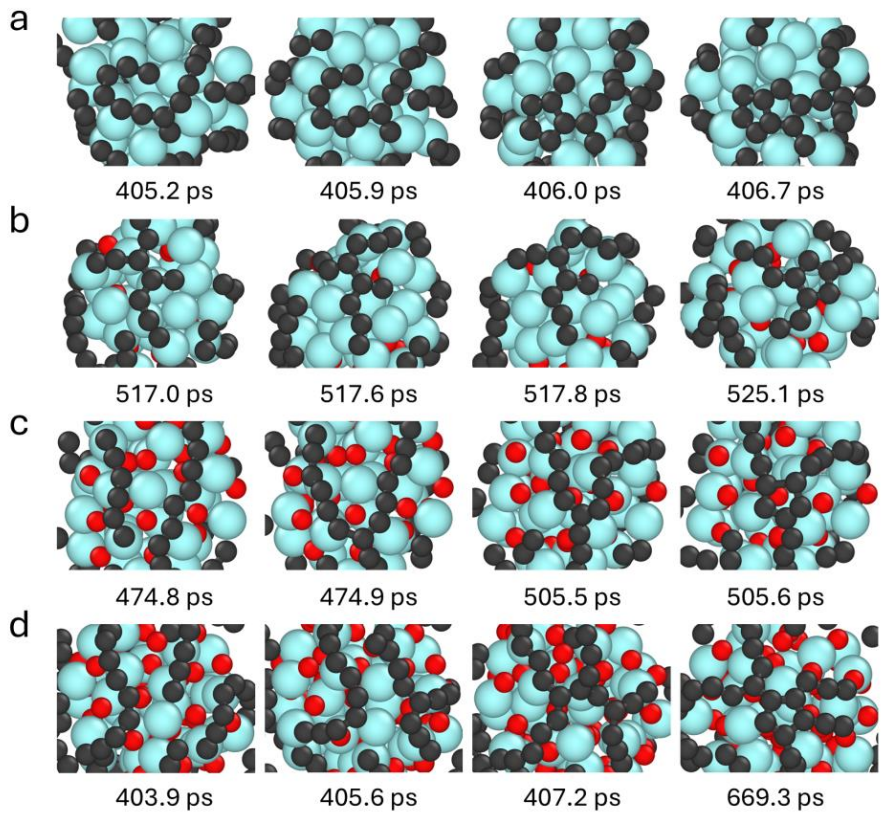

Figure 2. Exemplar pentagon-first ring nucleation mechanisms observed on (a) Fe, (b) $Fe_{38}O_{10}$, (c) $Fe_{38}O_{30}$ and (d) $Fe_{38}O_{50}$. Cyan, red and black spheres represent Fe, oxygen and C atoms, respectively.

### 3.3 How does oxygen influence the electronic structure of iron?

We conclude our discussion by proposing a new mechanism to account for the results discussed above. To begin, we reiterate that the main influence of oxygen on the SWCNT nucleation mechanism is the production of more active carbon growth precursors on the nanoparticle surface, increasing with higher oxygen content in the catalyst nanoparticle. This increase in active carbon species drives faster chain growth kinetics and impedes ring condensation, due to seemingly stronger Fe-C and weaker C-C bonding.

Analysis of the nanoparticle electronic structure during our CVD simulations, shown in Figure 3, reveals a new explanation that unifies these observations. Figure 3a presents the catalyst nanoparticle Fermi level observed during CVD as a function of the nanoparticle oxygen content. A clear trend is evident from this figure; upon increasing the oxygen content in the $Fe_{38}$ nanoparticle the Fermi level increases due to the oxidation of Fe via the partially occupied 3d valence shell by O 2p states. The increase in Fe 3d PDOS around the Fermi level is shown in Figure 3b. A similar increase that is proportional to oxygen content is shown in Figure 3c for O 2p PDOS. A maximum in the Fermi level is reached for the $Fe_{38}O_{40}$ model system used here (~1.5 eV higher in energy, compared to $Fe_{38}$), for which the catalyst's Fe:O stoichiometric ratio is essentially at parity. For oxygen loading beyond this point (i.e. $Fe_{38}O_{50}$), Figure 3a shows that the Fermi level decreases once again by ~0.5 eV. This is due to the increased O 2p PDOS at the Fermi level following broadening and enhanced hybridization of O 2p with Fe 3d states, which is absent at lower oxygen loadings (Figure 3c and Figure S4). On the basis of Figures 3a-c, the trends regarding the SWCNT nucleation mechanism discussed above can be explained using established heuristic concepts used in heterogeneous catalysis, such as d-band theory.[41, 42] This mechanism is depicted schematically in Figure 3d; by increasing the Fermi level of the catalyst nanoparticle, oxygen increases the likelihood of electron donation from Fe 3d states into unoccupied C 2p states of surface-adsorbed carbon moieties (e.g. C-C π* bonds etc.), as the energies of these orbitals are fixed, and become more commensurate with the nanoparticle Fermi level. Consequently, C-C bonds are weakened indirectly in the presence of oxygen, and this promotes the formation of more active carbon species on the nanoparticle surface, which in turn can increase the rate of chain growth, while impeding the condensation of rings. We note this mechanism has recently been used to explain indirect catalytic effects of Ru loading in CoRu nanoparticles.[REF]

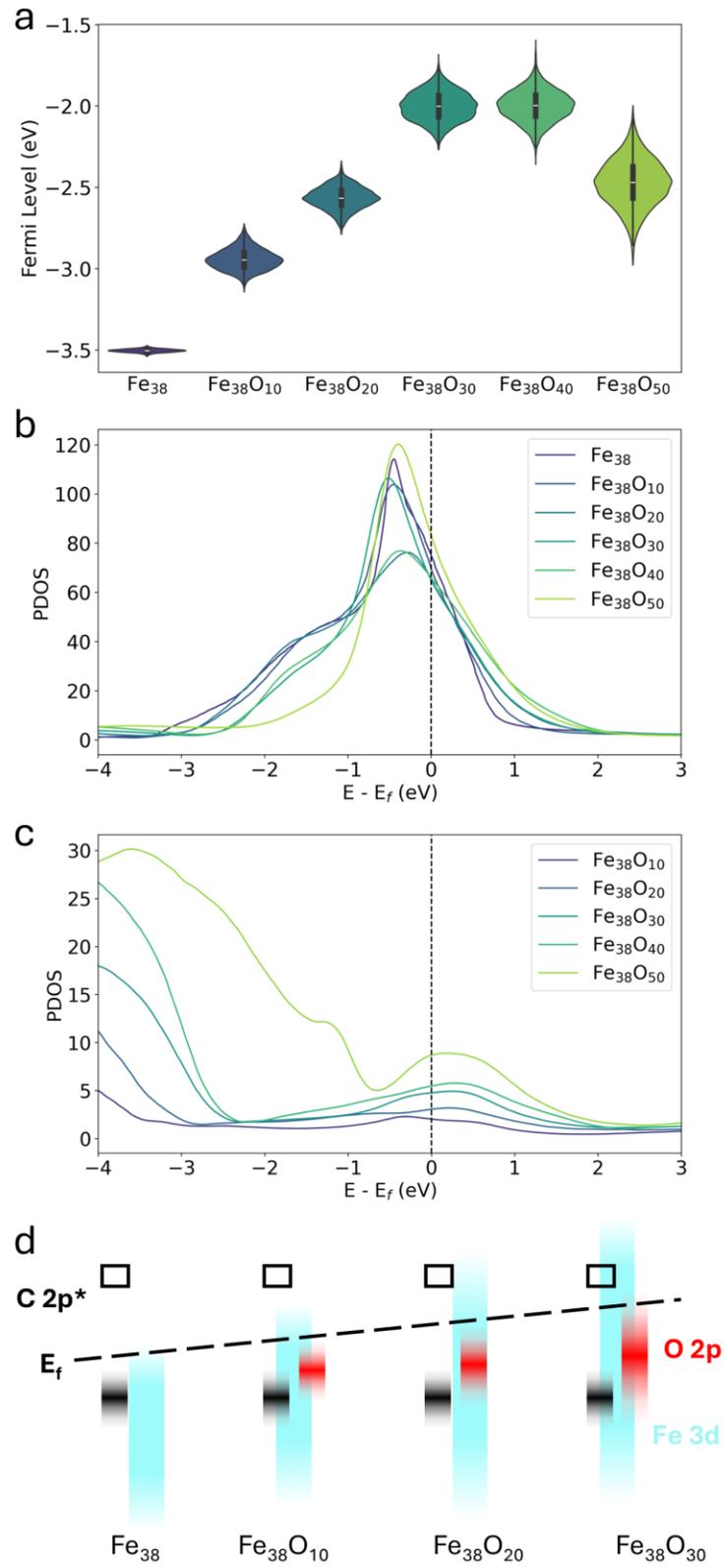

**Figure 3.** The Fermi level ($E_f$) of the $Fe_{38}$ catalyst nanoparticle is modulated by oxygen loading in as predicted by d-band theory.[41, 42] (a) Fermi level distributions of $Fe_{38}$ and $Fe_{38}O_x$ catalyst nanoparticles observed during simulated SWCNT nucleation. (b,c) Partial density of states (PDOS) for (b) Fe 3d states and (c) O 2p states near the Fermi level. (d) Schematic illustration of oxygen's modulation of Fe catalyst nanoparticle electronic structure, responsible for the observed changes in nucleation kinetics.

**Conclusion**

In this work, we employed quantum chemical simulations to investigate how oxygen influences the catalytic activity of iron during SWCNT nucleation. SCC-DFTB MD simulations demonstrate that increasing oxygen loading in the catalyst leads to more active carbon species on the surface, exhibiting stronger Fe-C and weaker C-C bonds with higher oxygen content. This leads to enhanced chain growth kinetics during annealing but impeded ring condensation at high oxygen loading. Notably, we attribute this phenomenon to changes in the electronic structure of the catalyst. The Fermi level of the catalyst increases proportionally with oxygen content, reaching a maximum at an approximately equal iron to oxygen ratio. This coincides with proportional increases in Fe 3d and O 2p states near the Fermi level. The elevated Fermi level promotes donation from Fe 3d orbitals into unoccupied C 2p orbitals of surface-bound carbon species, facilitating C-C bond formation and chain growth. This new insight into how oxygen influences the catalytic activity of iron during SWCNT nucleation by tuning the electronic structure of the catalyst, could unlock pathways for optimizing the degree of catalyst oxidation for diameter-controlled SWCNTs.

**ASSOCIATED CONTENT**

**Supporting Information**.

Additional data from DFTB MD simulations: bond counts, structures, partial atomic charges, particle Lindemann Index and PDOS.

**AUTHOR INFORMATION**

**Corresponding Author**

Ben McLean − School of Engineering, RMIT University, Victoria, 3001, Australia

orcid.org/0000-0001-7763-8828

Email: ben.mclean2@rmit.edu.au

**Author Contributions**

**Ben McLean:** Conceptualization, Methodology, Software, Validation, Formal Analysis, Investigation, Data Curation, Writing – Original Draft, Writing – Review & Editing, Visualization, Project Administration.

**Alister Page:** Investigation, Data Curation, Formal Analysis, Visualization, Writing – Review & Editing, Funding Acquisition.

**Feng Ding:** Resources, Writing – Review & Editing, Project Administration, Funding Acquisition.

**FUNDING SOURCES**

B.M. and F.D. acknowledge support from the Institute for Basic Science (IBS-R019-D1) of South Korea. A.J.P. acknowledges Australian Research Council funding (ARC DP210100873). This research was undertaken with the assistance of resources provided at the NCI National Facility systems at the Australian National University, through the National Computational Merit Allocation Scheme supported by the Australian Government.

**NOTES**

The authors declare no competing financial interests.

**ACKNOWLEDGMENT**